\documentclass[]{aa} % final version (more compact)
\usepackage{graphicx,epsfig}

\def\note[#1]{\vskip 0.25cm \noindent
              {\bf \underbar{Note}: #1 } \vskip 0.25cm}

\def\etal{et al.}
\def\eg{{\it e.g.}}
\def\teff{$T_{\rm eff}$}

\begin{document}
   \title{Detailed theoretical models for extra-solar planet-host stars}
   \titlerunning{Models for extra-solar planet-host stars}
   \subtitle{The ``red stragglers" HD\,37124 and HD\,46375}

   \author{Jo\~ao Fernandes\inst{1,2}
           \and
           Nuno C. Santos\inst{3,4}
          }

%   \offprints{J. Fernandes}

   \institute{
      Grupo de Astrof\'\i sica da Universidade de Coimbra, 
      Observat\'orio Astron\'omico da Universidade de Coimbra, Santa Clara, 3000 Coimbra, Portugal
      \and
      Departamento de Matem\'atica,
      FCTUC, Largo D. Dinis, Portugal\\
      \email{jmfernan@mat.uc.pt}
      \and
      Centro de Astronomia e Astrof\'\i sica da Universidade de Lisboa, Observat\'orio 
      Astron\'omico de Lisboa, Tapada da Ajuda, Lisboa, Portugal
      \and
      Observatoire de Gen\`eve, 51 ch. Des Maillettes, CH-1290, Sauverny, Switzerland\\
      \email{nuno.santos@oal.ul.pt}
             }

   \date{Received; accepted }

%- - - - - - - - - - - - - - - - - - - - - - - - - - - - - - - - - - -
   \abstract{In this paper we analyse and discuss the HR Diagram position of two extra-solar 
planet-host stars - HD\,37124 and HD\,46375 - by means of theoretical stellar evolution models. 
This work was triggered by the results obtained by Laws \etal\ (\cite{laws03}) who found 
that these stars were in contradiction to the expectation based on their high metallicity.\\ 
Fixing the age of both stars with the value based on their chromospheric activity levels 
and computing our own evolutionary models using the CESAM code, we are able to reproduce 
the observed luminosity, effective temperature and metallicity of both stars 
for a set of stellar parameters that are astrophysically reliable {even
if it is non-trivial to interpret the absolute values for these parameters.} 
Our results are discussed in the context of the stellar properties of low mass stars. 
}

   \maketitle
%
%_____________________________________________________________________
\section{Introduction}
During the last few years a number of different detailed spectroscopic studies have 
revealed that planet-host stars are (on average) very metal-rich compared to field dwarfs (e.g. Gonzalez \etal\ \cite{Gonzalez97,Gonzalez01}; 
Santos \etal\ \cite{santos00,santos01,santos04}; Laws \etal\ \cite{laws03}).
Current results favour that this ``excess'' metallicity arises from the cloud of
gas and dust that generated the star/planetary-system, suggesting 
that the formation of giant planets (at least the kind currently beeing discovered) is 
critically dependent on the grain content of the proto-planetary disk 
(e.g. Santos \etal\ \cite{santos04}). This result has an enormous impact 
on theories of planetary formation.

In one of the studies mentioned above, Laws \etal\ (\cite{laws03}) analysed a sample 
of 30 F, G, K stars with companions (planets or brown dwarfs). They preformed spectroscopic 
observations in order to determinate their atmospheric parameters, including the 
effective temperature and metallicity. 
All the stars had accurate {\it HIPPARCOS} parallaxes, and thus the stellar position in the 
Hertzsprung-Russel Diagram (HRD) could be obtained 
precisely enough to derive stellar masses and ages from theoretical stellar models. 
To derive these parameters, Laws \etal\ (\cite{laws03}) used the evolutionary 
tracks from the Padova group (Salasnich \etal\ \cite{salasnich00}). In their conclusions 
it is pointed out that for four stars in their sample (\object{HD\,6434}, \object{HD\,37124}, 
\object{HD\,46375}, \object{HD\,168746}) the derived ages 
were above 14.8 Gy in spite of being Population I stars. They called this quartet 
the ``red stragglers'', because of their unexpected red position in the HRD. 

The explanation for the position of these four stars in the HRD is not clear. 
Laws \etal\ (\cite{laws03}) suggested that it could be related to the presence of close-in 
low-mass stellar companions. However, to our knowledge none has been found by any Adaptative 
Optics survey. Laws \etal\ (\cite{laws03}) ``recommend that these stars be the 
targets of detailed studies to determine the source(s) of their aberrant physical characteristics". 
We perform this study starting from a detailed analysis 
of the position of two of these objects in the HRD, to verify to what extent 
these two objects can be explained by current models of stellar evolution.

The knowledge of the HRD position of a 
star is dependent on many different variables. 
To first order, it is fixed by the stellar mass ($M$), the initial individual 
abundance of helium ($Y$) and metals ($Z$), and age ($t_{\star}$). The resolution of the equations 
of internal structure using the above parameters will give the values for the surface 
temperature and luminosity. However, the physical inputs chosen to describe 
the stellar interior also constrain the evolution of the star in the HRD. In particular, some 
mechanisms insufficiently known such as the convection, rotation and diffusion are dependent on 
free parameters. For instance, in the framework of the Mixing Length Theory (MLT, see below) 
currently used to model stellar convection, one more (unknown) variable has to be considered: 
the mixing-length parameter ($\alpha$). So, to model a star correctly, we must determine the 
five parameters. 

From the observational point of view, a detailed spectroscopic analysis of a single star may allows us 
to estimate its metallicity ($[Fe/H]$ or $Z/X$, where $X = 1-Y-Z$ is the abundance of hydrogen), 
its luminosity ($L$)\footnote{Better determined if the distance to
the star is known, as it is in our case.} and its effective temperature, \teff. 
We are thus left with three known observables. In other words, we have an indeterminate problem: 
five unknowns to three observables. Gravity is also an 
issue in the spectroscopic analysis. So, a determination of the mass should, in principle, also be possible. However, a typical realistic error in $\log g$ is 
usually above 0.1\,dex (\eg\ Laws \etal\ \cite{laws03}). 
This error implies an uncertainty on the derived solar mass stars of more 
than 0.15 $M_\odot$. Clearly, this error is too large to allow the gravity to be a real mass constraint.

This indeterminate problem can be solved for the Sun, for which the mass, metallicity and
age are observationally known (Christensen-Dalsgaard \cite{jcd82})
and for some binary systems where individual masses are available
(\eg\ Noels \etal\ \cite{noels91}; Fernandes \etal\ \cite{fernandes98}). 
But for a single star (without a close-in stellar companion), other than the Sun, commonly 
there are more unknowns than observables, the situation that we have here.

In this work we analyze in detail two of the ``red-stragglers'' 
discussed by Laws \etal\ (\cite{laws03}), namely \object{HD\,37124} and \object{HD\,46375}. 
Our goal is to find physically consistent models that are able to explain the position of 
these stars in the HRD. We have excluded \object{HD\,168746} 
and \object{HD\,6434} from this study: the former because it 
has no available chromospheric activity age (which would give us a further constraint -- see below), 
and the latter because with a $[Fe/H]=-0.55\pm0.07$ 
it is on the frontier between Population\,I and Population\,II stars. In the modelling of Population\,II 
stars some effects become important. This is the case of the diffusion of chemical 
elements, NLTE corrections for the metallicity and $\alpha$-elements contribution to the opacity 
(Lebreton \etal\ \cite{lebreton99}). So, as the cumulated effects can be a supplementary 
source of errors, we prefer to exclude this case.

In Sec.\,2 we present the stellar models and the theoretical description of our methodology 
used for modelling the stars. In Sec.\,3 we compute models for \object{HD\,37124} 
and \object{HD\,46375}. In Sec.\,4 we present a discussion and conclusions in the context of 
the stellar proprieties of low mass stars. In particular, we show that the two stars studied
in the current paper can be correctly modelled for some physically reliable values of the
stellar parameters.

%_____________________________________________________________________

\section{Input physics in the theoretical stellar models and modelling method}

We compute our models specially for the stars studied in this work.
The stellar evolution calculations were done with the CESAM code
version\,3 (Morel \cite{morel97}), running in the Coimbra Observatory, in Portugal.

Details on the physics of these models can be found in Lebreton \etal\ (\cite{lebreton99}):
the CEFF equation of state, including Coulomb corrections
to the pressure (Eggleton \etal\ \cite{eggleton73}, Christensen-Dalsgaard \cite{jcd91});
nuclear reactions rates given by Caughlan and Fowler (\cite{caughlan88});
solar mixture from Grevesse and Noels (\cite{grevesse93});
OPAL opacities (Iglesias and Rogers \cite{iglesias96})
complemented at low temperatures by opacity data from
Alexander and Ferguson (\cite{alexander93}) following a 
prescription of Houdek and Rogl (\cite{houdek96}).
The atmosphere is described with an Eddington $T(\tau)$-law;
convection is treated according to the mixing-length 
theory from Bohm-Vitense (\cite{bohm-vitense58})
leaving the mixing-length ($\alpha\times H_p$) unknown,
and thus $\alpha$ as a free parameter, where $H_p$ is the pressure scale height.
With these input physics the solar model fits the observed luminosity 
and radius, with a relative precision of  $10^{-3}$, with $\alpha{=}1.63$, helium abundance $Y{=}0.268$ and $Z=0.0175$ for 
the common accepted solar age of 4.6~Gyr (Dziembowski \etal\ \cite{dziembowski99}) and the ratio of the 
solar mixture from Grevesse and Noels (\cite{grevesse93}). {This model does not take into account microscopic diffusion of helium or metals.}  

There are other mechanisms, not included in this work, 
that could affect the HRD position of a model, such as the rotation (Maeder and Meynet \cite{maeder00}), 
overshooting of the convective core, or helium and metal gravitational settling 
(microscopic diffusion). However, the stars is this work are slow rotators (as are all 
planet-hosts, since it is more difficult to obtain accurate radial-velocities for fast rotating stars) 
and solar mass stars do not develop permanent convective cores (Ribas \etal\ \cite{ribas00}). On the 
other hand, diffusion has a marginal effect on the HRD position for Population I solar mass stars. Lastennet \etal\ (\cite{lastennet03}) have shown that for a 0.98$M_\odot$ star with Z=0.012, the inclusion of 
helium and metal diffusion may change the \teff\ of the evolutionary track by no more than 65\,K, 
even for an old age of 10 Gy. For Population I stars, the density in external layers is 
sufficiently high to break the diffusion efficiency. We thus expect that the lack of these 
mechanisms do not considerably change our results. 

{As discussed above, modelling a single star, for a fixed age, is reduced to the problem 
of finding the solutions of the four stellar parameters $M$, $Y$, $Z$ and $\alpha$, that 
adjust the three observational parameters $L$, \teff, and $Z/X$.

Inspired by solar modelling 
(Christensen-Dalsgaard \cite{jcd82}), we construct the following system 
composed of the three equations: 

\begin{equation}
(\log\frac{L}{L_\odot})_{obs} = (\log\frac{L}{L_\odot})_{ref} + 
{{\sum^4_i}} (\frac{\partial {\log\frac{L}{L_\odot}}}{\partial X_i})_{j\neq i} 
\times (X_{i}- X_{i_{ref}})
\end{equation}

\begin{equation}                                                                
\log T_{{\rm eff}_{obs}} = \log T_{{\rm eff}_{ref}}+                            
{{\sum^4_i}} (\frac{\partial {\log T_{\rm eff}}}{\partial X_i})_{j\neq i}       
\times (X_{i}- X_{i_{ref}})$$                                                   
\end{equation}

\begin{equation}
({\frac{Z}{X}})_{obs} = \frac{Z}{1-Y-Z} 
\end{equation}
where $X_i$ with $i=1, 2, 3$, and~$4$ are 
$M$, $Y$, $Z$, $\alpha$, respectively. The subscript ``ref'' indicates the reference stellar evolutionary model that falls inside the observational error bars. 
We consider as a satisfactory solution the sample $(M,Y,Z,\alpha)$, obtained by solving 
the above system assuming the linearity of the variation of the $\log L/L_\odot$ and $\log ~$\teff\ with mass, 
helium, metals and $\alpha$ for a fixed value of the age. Theoretically, each solution of the system should fit exactly 
the observed luminosity and effective temperature, for the fixed age. In practice, thanks to our linear approximation, 
in some rare situations the solution is outside the error bars. However, in these situations the solution is still near and a small change in one parameter is enough to put the evolutionary track in the error bar.
On the other hand, as we have more unknowns than equations, the above system has an infinite number of solutions. 
In this work we chose to present five solutions for each star in order to illustrate the range of 
variation of the stellar parameters $(M,Y,Z,\alpha)$.
}

%_____________________________________________________________________

\section{Modelling HD 37124 and HD 46375}

In the following we will use the ages for \object{HD\,37124} and \object{HD\,46375} that
have been estimated by Laws \etal\ (\cite{laws03}) using a chromospheric activity-age 
calibration from Donahue (\cite{donahue93}). The knowledge of this age gives us the possibility to
limit the number of models available. We have thus fixed the age for each star 
(3.9 and for 4.5\,Gyr for \object{HD\,37124} and \object{HD\,46375}, respectively) and estimated 
the other four unknown parameters ($M$, $Y$, $Z$, $\alpha$) by theoretical stellar evolutionary 
models, taking into account the three observables ($L$,~\teff\ ,$Z/X$). {So, opposite to Laws \etal\ (\cite{laws03}), we have chosen not to fix the values of $\alpha$ and $\frac{\Delta Y}{\Delta Z}$ 
in our models}. We note that our major goal is to find an astrophysically reliable solution for the position of 
these stars in the HRD, i.e. a solution that matches the observable quantities. We thus prefer to
be as strict as possible in our observable constraints. Leaving the age as a free
parameter would, of course, make this solution easier to obtain, but this solution would
probably not be reliable.

\subsection{HD 37124}

\subsubsection{HD 37124: observations}

\object{HD\,37124} is a G4 IV-V star (V=7.68; B-V=0.667) in the solar neighbourhood at a distance
of 33.2 \,{\it pc} ({\it HIPPARCOS} -- Perryman \etal\ \cite{perryman97}). 
Vogt \etal\ (\cite{vogt00}) reported the discovery of a candidate planet around this star,
followed by the discovery of a second planetary-mass companion in the system (Butler et al. \cite{butler2003}). 
Since then, particular attention has been paid to this object, e.g., detailed spectroscopic analyses (Laws \etal\ \cite{laws03}; Santos \etal\ \cite{santos04}) 
or studies of the dynamical stability of the discovered planetary system (Zhou and Sun \cite{zhou03}).

In the following we will consider the spectroscopic determinations and magnitudes 
given by Laws \etal\ (\cite{laws03}): \teff\ =$5551\pm 34 K$; ${[Fe/H]=-0.37\pm 0.03}$ 
and $M_{V}~=5.07\pm 0.08$. These spectroscopic results are in excellent agreement with 
the recent values obtained by Santos \etal\ (\cite{santos04}): \teff\ =$5546\pm 30 K$; 
${[Fe/H]=-0.38\pm 0.04}$.

For Population\,I stars, the abundance of metals $Z$ is related to 
the $[Fe/H]$ by $[Fe/H] {=} \log (Z/X) {-} \log(Z/X)_{\odot}$,
where $(Z/X)_{\odot}{=}0.0245$ is the ratio of the solar 
mixture (Grevesse and Noels \cite{grevesse93}). This implies 
$(Z/X){=}0.0105{\pm}0.0008$ for \object{HD\,37124}.

Using the \teff\ of Laws \etal\ (\cite{laws03}) we derive a bolometric correction 
from Flower (\cite{flower96}) of $BC=-0.13\pm0.01$. This implies a bolometric 
magnitude of $M_{bol}=4.94\pm0.08$. Assuming that the bolometric magnitude 
of the Sun is 4.75 (e.g. Cayrel \cite{cayrel02}), we obtain a luminosity of $0.84{\pm}0.06$ 
times the solar luminosity ($L_\odot$) for \object{HD\,37124}. From the Stefan-Boltzmann law we can thus 
estimate the radius of this star: $0.99 \pm 0.05~R_\odot$. This result is in very good 
agreement with the infrared photometry determination of $1.004\pm0.039$ from 
Ribas \etal\ (\cite{ribas03}). The observable stellar parameters used to constraint 
the models are summarised in Table~\ref{tab:observacoes}.  

%....................................................................

\begin{table}
      \caption[]{Observations of HD\,37124 and HD\,46375 (see text for details).}
         \label{tab:observacoes}
     $$ 
         \begin{array}{cccc}
            \hline
            \noalign{\smallskip}
             star & L/L_{\odot} & Teff~(K)& Z/X
           				 \\
            \noalign{\smallskip}
            \hline
            \noalign{\smallskip}
            HD\,37124 &0.84\pm 0.06 & 5551\pm 34 & 0.0105\pm 0.0008  \\
HD\,46375&0.74\pm 0.06 & 5241\pm 44 & 0.049\pm 0.004  \\
            \noalign{\smallskip}
            \hline
         \end{array}
     $$ 
   \end{table}

%......................................................................

\subsubsection{HD 37124: models}

We have computed several models for \object{HD\,37124}, fixing the age 
at 3.9 Gy from Laws \etal\ (\cite{laws03}), in the following range of parameters: 
4 values of mass in $[0.75,0.95]$, 4 values of helium in $[0.24,0.30]$, 4 values of metallicity 
in $[0.00669,0.01000]$ and 5 values of $\alpha$ in $[0.8,1.6]$. 
This sample of models allows us to obtain the derivatives of $L$ and \teff\ in relation 
to ($M$, $Y$, $Z$, $\alpha$) in the system of equations (1), (2) and (3). Using the observed values ($L$, \teff, $Z/X$) we can then establish a linear system for 
the reference model with the following characteristics: $M{=}0.94~M_\odot$, $Y{=}0.24$, 
$Z{=}0.0079$, and $\alpha{=}1.00$. This model reproduces the observed quantities 
($L$, \teff\ , {Z/X}), within the errors, for an age of 3.9~Gyr. 

\begin{figure}[t]
\epsfig{width=\hsize,file=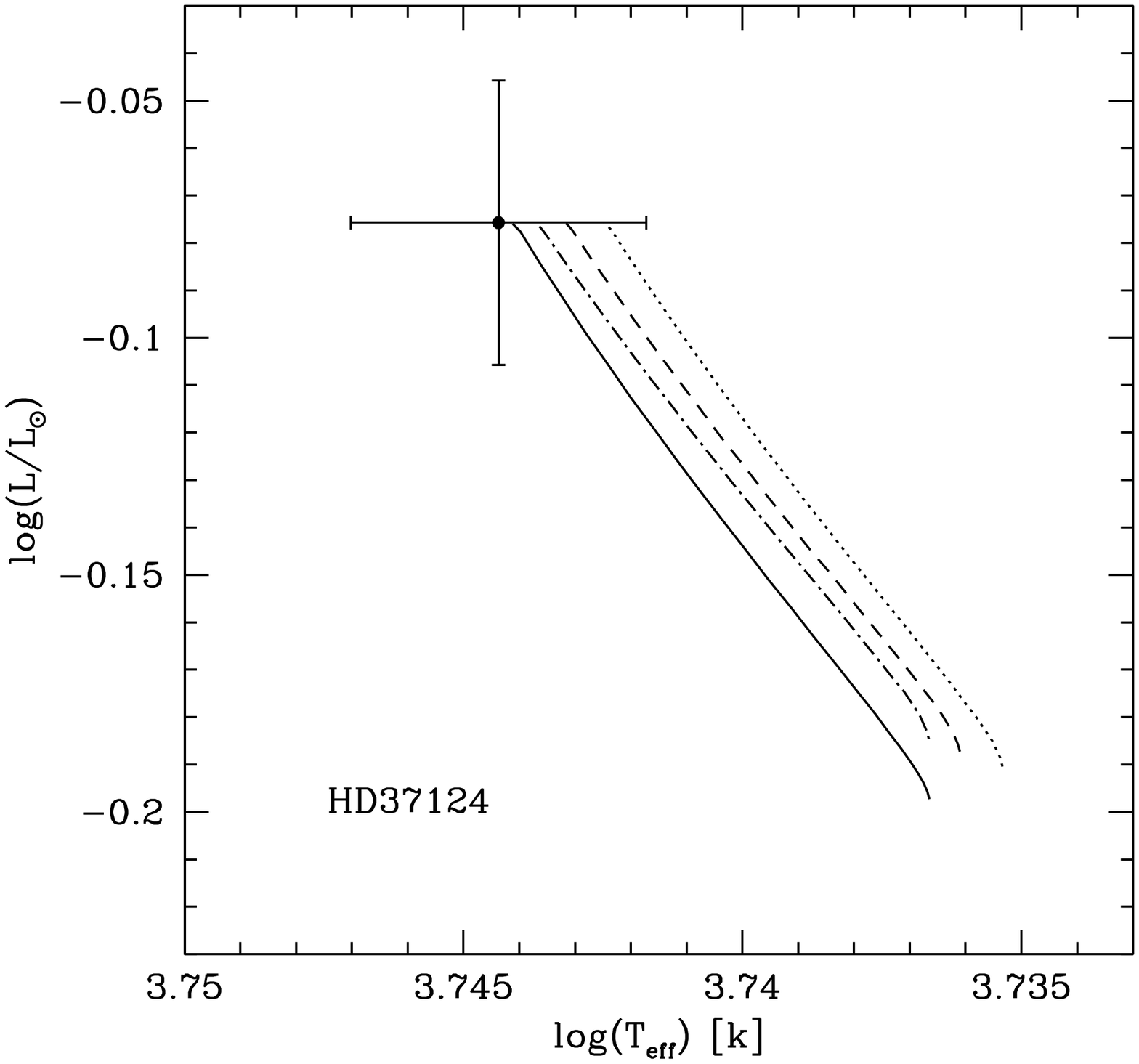}
\caption[]{Evolutionary tracks for HD\,37124. The different lines represent models
1 (solid line), 2 (dotted line), 3 (dashed line), and 4 (dot-dashed line). See text for more details.}
\label{fig:test-evol}
\end{figure}

In Table~\ref{tab:models-hr} we present the models from 1 to 4 that reproduce the observations within 
the error bars and in Fig.~\ref{fig:test-evol} we provide the corresponding evolutionary tracks 
in the HRD from the ZAMS to 3.9 Gy. 
The number of possible models that fall inside 
the observed error bars is infinite. The first four models were chosen in order to show the dependence 
of the solution on $M$, $Y$, $Z$ and $\alpha$. {Model 5 is
computed taking into account
an increase in the observed \teff\ of about $3\sigma$
(see Discussion and conclusions). The correponding luminosity
is 0.82$L_\odot$, still inside the original error bar}.

%....................................................................
\begin{table}
      \caption[]{Models of HR 37124 for an age of 3.9 Gy.
      $M$ is the mass of the model (in $M_\odot$),
      $Y$ is the initial helium abundance, $Z$ is the initial metal abundance,
      $\alpha$ the mixing length (in units of the local pressure scale height).}
         \label{tab:models-hr}
     $$ 
         \begin{array}{lcccccc}
            \hline
            \noalign{\smallskip}
            {\rm Model} \qquad & M & Y & Z &\alpha & T_{\rm eff}~(K)& L/L_\odot
           				 \\
            \noalign{\smallskip}
            \hline
            \noalign{\smallskip}
            model~1 & 0.86 & 0.29 & 0.0074 & 0.85 & 5548 & 0.84\\
            model~2 & 0.90 & 0.27 & 0.0076 & 0.90 & 5526 & 0.84\\
            model~3 & 0.92 & 0.25 & 0.0079 & 0.95 & 5535 & 0.84\\
            model~4 & 0.94 & 0.24 & 0.0079 & 1.00 & 5542 & 0.84 \\
	    model~5 & 0.90 & 0.27 & 0.0076 & 1.10 & 5668 & 0.84 \\
            \noalign{\smallskip}
            \hline
         \end{array}
     $$ 
   \end{table}

%......................................................................

The five solutions have in common low values of the mixing-length parameter, $\alpha$. This 
is probably the reason why Laws \etal\ (\cite{laws03}) did not find a satisfactory solution in their 
analysis\footnote{By "satisfactory solution" we mean the one that, for a
fixed age, reproduces the observational constraints (L, \teff, Z/X).}. 
In fact, all the Padova isochrones are built using the solar $\alpha$ value ($\sim 1.7$), the same for all stellar models. This approximation, currently made on the isochrone computations for {simplicity}, is not clear: why should stars with different masses, chemical composition and evolutionary status, have the same $\alpha$ ? Some recent results on stellar systems conclude that there are variations of $\alpha$ with mass (Lebreton \etal\ \cite{lebreton02}), including values of $\alpha$ lower than 1.0 (Lastennet \etal\ \cite{lastennet03}). {On the other hand, the helium and mass are values expected for this star.}

\subsection{HD 46375}

\subsubsection{HD 46375: observations}

\object{HD\,46475} is a K1 IV-V star (V=7.91; B-V=0.860), lying in the solar 
neighbourhood: $d = 33.4~pc$ (Perryman \cite{perryman97}). This star was 
recently found to be orbited by a possible planetary-mass companion 
(Marcy \etal\ \cite{marcy00}).

To build the observed HRD we consider here the parameters derived by 
Laws \etal\ (\cite{laws03}): \teff\ =$5241\pm 44 K$; ${[Fe/H]=+0.30\pm0.03}$ and $M_{V}~=5.29\pm 0.08$. 
These spectroscopic results are in good agreement with the recent results by Santos \etal\ (\cite{santos04}) 
for the effective temperature (\teff\ =$5268\pm55\,K$) and are in fair agreement with the 
metallicity (${[Fe/H]=+0.20\pm0.06}$). Computed as described above, the metal abundance of this 
star is $(Z/X){=}0.049{\pm}0.004$.

Using the \teff\ of Laws \etal\ (\cite{laws03}) we can derive the bolometric correction 
of $BC=-0.21\pm0.03$ (Flower \cite{flower96}), a value that implies a bolometric magnitude 
$M_{bol}=5.08\pm0.09$, and a luminosity of $0.74{\pm}0.06~L_\odot$; the corresponding 
stellar radius is $1.05\pm0.06~R_\odot$, once more in very good agreement with 
Ribas \etal\ (\cite{ribas03}): $1.005\pm0.036 R_\odot$. The observations chosen 
to constrain the models are provided in Table~\ref{tab:observacoes}.

\subsubsection{HD 46375: models}

As done for \object{HD\,37124}, we computed several models for \object{HD\,46375}, 
fixing the age at 4.5\,Gyr from Laws \etal\ (\cite{laws03}) in the following range of stellar parameters: 
4 values of mass in $[0.95,1.05]$, 4 values of helium in $[0.24,0.30]$, 
4 values of metals in $[0.025,0.035]$ and 4 values of $\alpha$~in~$[0.8,1.6]$. 

The reference model has the following chacteristics: $M{=}1.00~M_\odot$, $Y{=}0.30$, $Z{=}0.036$ 
and $\alpha{=}1.0$. This model is sufficiently near the observed quantities 
($L$, \teff\ , {Z/X}) for an age of 4.5~Gyr. 

\begin{figure}[t]
\epsfig{width=\hsize,file=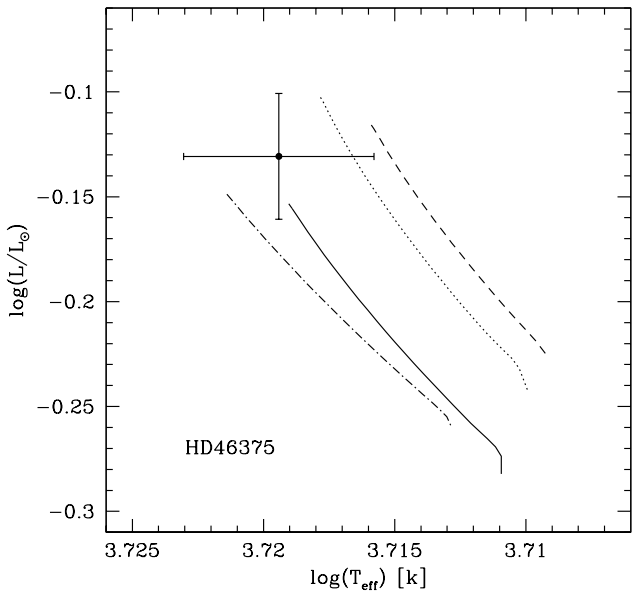}
\caption[]{Evolutionary tracks for HD\,46375. The different lines represent models
6 (solid line), 7 (dotted line), 8 (dashed line), and 9 (dot-dashed line). See text for more details.}
\label{fig:test-evol1}
\end{figure}

In Table~\ref{tab:models-hr1} we present four derived models, from 6 to 9, that reproduce the observations within the 
error bars and in Fig.~\ref{fig:test-evol1} we plot the correspondent evolutionary tracks in 
the HRD from the ZAMS to 4.5 Gyr. As for the previous star we chose four models to 
give the range of variation of each parameter $M$, $Y$, $Z$ and $\alpha$. 
{Model 10 is computed taking into account 
an increase of $3\sigma$ in the observed \teff\ . 
As previously, the corresponding luminosity is 0.71$L_\odot$, falling 
inside the original error bars.}

%......................................................................

\begin{table}
      \caption[]{Models of HD 46375, for an age of 4.5 Gy.
      $M$ is the mass of the model (in $M_\odot$),
      $Y$ is the initial helium abundance, $Z$ is the initial metal abundance,
      $\alpha$ the mixing length (in units of the local pressure scale height).}
         \label{tab:models-hr1}
     $$ 
         \begin{array}{lcccccc}
            \hline
            \noalign{\smallskip}
            {\rm Model} \qquad & M & Y & Z &\alpha & T_{\rm eff}~(K)& L/L_\odot
           				 \\
            \noalign{\smallskip}
            \hline
            \noalign{\smallskip}
            model~6 & 0.96 & 0.31 &  0.032 & 1.15  & 5236& 0.71 \\
            model~7 &  0.98& 0.31 & 0.032 & 1.05 &5224 &0.79 \\
            model~8 & 1.01 & 0.29 & 0.033 & 1.10 & 5200& 0.77 \\
	    model~9 & 1.03 & 0.27 & 0.034 & 1.35 & 5272& 0.71 \\
	    model~10 & 1.03 & 0.27 & 0.034 & 1.55 & 5369 & 0.72 \\
            \noalign{\smallskip}
            \hline
         \end{array}
     $$ 
   \end{table}

%......................................................................

{As for HD\,37124, the models from 6 to 9 present low $\alpha$ values confirming that a probable explanation for the difficulty of 
Laws \etal\ (\cite{laws03}) in finding a solution with a "reliable" age can be attributed to the constant value of $\alpha$ in the Padova models.}       

\section{Discussion and conclusions}

In this paper we present detailed evolutionary models that reproduce the observed HRD position and 
metallicity of the planet-host stars \object{HD\,37124} and \object{HD\,46375}. In order to 
reduce the high degeneracy degree of modelling a single star we fixed the stellar age given 
by the age-cromospheric activity relation.

The main goal of this paper is to show that it is possible to model these stars using astrophysically 
reliable values of mass, helium and metal abundances, and mixing length parameters. We 
found that this is indeed the case, although $\alpha$ and $\frac{\Delta Y}{\Delta Z}$ values are not considered 
as a constants from one model to another. The hypothesis of constant values is current in published isochrones. We propose that the ``aberrant" solutions found 
by Laws \etal\ (\cite{laws03}) for \object{HD\,37124} and \object{HD\,46375} are caused by the 
assumption of the above hypothesis, particularly the one about the mixing length parameter.

Eventual differences 
between the internal structure of the CESAM and Padova evolutionary codes, as well as 
between our modelling method and the one used by Laws \etal\ (\cite{laws03}), 
can't explain our results. The three following reasons support this: first, the input physics, namely the radiative opacities, equation of state and convection formalism, are similar in the two codes; second, we confirm the difficulty of Laws \etal\ (\cite{laws03}) in modeling the above stars using solar values for $\alpha$ and $\frac{\Delta Y}{\Delta Z}$; and third, we tested our modelling method in \object{HD\,117176}, one star also studied by Laws \etal\ (\cite{laws03}), and for which they have 
found a satisfactory solution. Using the mass derived by these authors, as well as the same $\alpha$ and $\frac{\Delta Y}{\Delta Z}$ used in the Padova isochrones, we find a solution in the HRD 
for an age of 8.8 Gy, close to the 7.9\,Gyr derived by them.

Among an infinite number of combinations of stellar parameters ($M$, $Y$, $Z$, $\alpha$) that, 
for a fixed age, reproduce the observables ($L$, \teff\ , $Z/X$)\footnote{As mentioned above, the 
modelling of a single star other than the Sun is an indeterminate problem, given that there are 
more unknowns than observables.} we have computed four models for each star, inside an acceptable 
range of variation for each parameter. As we do not have complementary observational constraints 
we cannot choose the best model(s). However, some astrophysical arguments, such as 
the reliable values of the helium-to-metals chemical enrichment parameter $\frac{\Delta Y}{\Delta Z}$ (\eg\ Jimenez \etal\ \cite{jimenez03}) 
could allow us to constrain the solutions. {Note that our solutions are age dependent}. 

Understanding why these stars are well modelled using 
a low $\alpha$ value, while that is not the case for other stars in the sample of Laws, 
is beyond the scope of this paper. The MLT is a local, 1D and non-compressive theory to explain 
a non-local, 3D and highly compressible system, as is stellar convection. Many works and authors (e.g. Canuto and Mazzitelli \cite{canuto91}; Ludwig \etal\ \cite{ludwig99}), 
have already shown the limitations of this formalism. However MLT is easy to implement in stellar evolutionary 
codes and is a relatively successful in reproducing global observations of the Sun and stellar systems. 
The low values of $\alpha$ found in this work, probably, reflect the limitations on the MLT itself 
more than a real physical effect. The analysis of these stars using alternative convective description such as the 
formalism of Canuto and Mazzitelli (\cite{canuto91}) could give important constraints on 
the super-adiabatic layer in the convective region, where convective efficiency is strongly dependent 
on $\alpha$. However, we cannot exclude that the derived low values for $\alpha$ are due to some other 
unknown effect. 

It is well known that the \teff\ of low mass stellar models depends on $\alpha$, thanks to its influence in the super-adiabatic layer and, thus, on stellar radius. Are there erroneous \teff\ determinations? 
{We have checked that if the observed \teff\ is increased by 
3$\sigma$ for each star, we can find a solution with a 
high $\alpha$ value for both stars (models 5 and 10). In particular,  
model 10 has an $\alpha$ value closer to solar. On the other hand these 
changes in \teff\ must be weighted by the fact that systematic 
differences between various observational temperature scales are 
never above 100-200\,K (c.f. Laws \etal\ \cite{laws03}; 
Santos \etal\ \cite{santos04}). So at least for \object{HD\,46375} the 
reason for the ``abberant solution'' found
by Laws \etal\ (\cite{laws03}) may be an underestimation of \teff\ .}

{The solutions found for each star are all in
a range lower than 0.1 solar masses}. This can be particularly important for the estimation of the mass the planets harboring stars,
and that could have a strong impact e.g. on the study of the relation between the
stellar mass and the frequency of giant planets (Santos et al. \cite{santos2003}; Laws 
et al. \cite{laws03}). Furthermore, the wrong masses
derived from the standard evolutionary models (with, among other things, $\alpha$ fixed to
the solar value) may imply incorrect derivations of the surface gravities using Hipparcos parallaxes (e.g. 
Nissen et al. \cite{nissen1997}). If the values of $\alpha$ depend e.g. on the effective temperature
of the star, this problem may induce a systematic error in the gravities with temperature when comparing spectroscopic
and parallax-based $\log{g}$ values, an
effect that was mentioned in Santos et al. (\cite{santos04}).

If indeed parameters like $\alpha$ vary from star to star, the determination
of accurate ages by means of isochrone fitting may introduce important systematic errors.
This was the case for the two stars studied in this paper, but may be the case
for many other objects {thus, specific stellar models must be computed for individual objects}.

%_____________________________________________________________________

\begin{acknowledgements}
We would like to thank the referee, Dr. Achim Weiss, for his
helpful suggestions to improve the manuscrit. This work was supported in part by Funda\c{c}\~ao para a Ci\^encia e a Tecnologia - Portugal,
through {POCTI-SFA-2-675}. JF gratefully thank David Valls-Gabaud 
who installed the CESAM code at the Coimbra Observatory. 
This work was done with the help of the CDS-Strasbourg Data Base.
Support from Funda\c{c}\~ao para a Ci\^encia e Tecnologia (Portugal) to N.C.S. in the form of a scholarship is gratefully acknowledged.
\end{acknowledgements}

%_____________________________________________________________________

%_____________________________________________________________________
\end{document}